\documentclass{article}

\usepackage{PRIMEarxiv}

\usepackage[utf8]{inputenc} 
\usepackage[T1]{fontenc}    
\usepackage{hyperref}       
\usepackage{url}            
\usepackage{booktabs}       
\usepackage{amsfonts}       
\usepackage{nicefrac}       
\usepackage{microtype}      
\usepackage{lipsum}
\usepackage{fancyhdr}       
\usepackage{graphicx}       
\graphicspath{{media/}}     

\usepackage{amsmath}
\usepackage{subcaption}
\usepackage[dvipsnames]{xcolor}

\newcommand{\methodname}{\textit{RED-DOT}}

\title{RED-DOT: Multimodal Fact-checking via Relevant Evidence Detection
}

\usepackage{authblk}

\author[1, 2]{\small Stefanos-Iordanis Papadopoulos\thanks{Corresponding author} \ }
\author[1]{\small Christos Koutlis}
\author[1]{\small Symeon Papadopoulos}
\author[2]{\small Panagiotis C. Petrantonakis}

\affil[1]{\footnotesize Information Technology Institute, Centre for Research \& Technology, Hellas.}
\affil[2]{\footnotesize Department of Electrical \& Computer Engineering, Aristotle University of Thessaloniki.}
\affil[ ]{\textit {\{stefpapad,ckoutlis,papadop\}@iti.gr, \textit{ppetrant@ece.auth.gr}}}

\begin{document}
\maketitle

\begin{abstract}

Online misinformation is often multimodal in nature, i.e., it is caused by misleading associations between texts and accompanying images. 
To support the fact-checking process, researchers have been recently developing automatic multimodal methods that gather and analyze external information, \textit{evidence}, related to the image-text pairs under examination. 
However, prior works assumed all external information collected from the web to be relevant.  
In this study, we introduce a ``Relevant Evidence Detection'' (RED) module to discern whether each piece of evidence is relevant, to support or refute the claim. 
Specifically, we develop the ``Relevant Evidence Detection Directed Transformer'' (\methodname) and explore multiple architectural variants (e.g., single or dual-stage) and mechanisms (e.g., ``guided attention''). Extensive ablation and comparative experiments demonstrate that \methodname \ achieves significant improvements over the state-of-the-art (SotA) on the VERITE benchmark by up to 33.7\%.
Furthermore, our evidence re-ranking and element-wise modality fusion led to \methodname \ surpassing the SotA on NewsCLIPings+ by up to 3\% without the need for numerous evidence or multiple backbone encoders. 
We release our code at: \url{https://github.com/stevejpapad/relevant-evidence-detection}.

\end{abstract}


\section{Introduction}
\label{sec:intro}

\footnotetext[1]{This work has been submitted to the IEEE for possible publication. Copyright may be transferred without notice, after which this version may no longer be accessible.}

\begin{figure*}
    \centering
    \includegraphics[width=\textwidth]{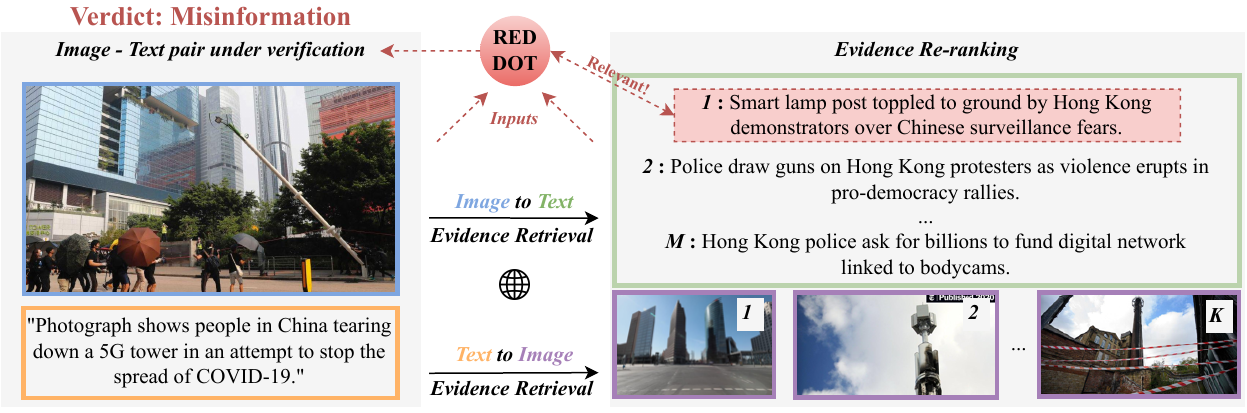}
    \caption{\textcolor{RoyalBlue}{\textbf{Image}}-\textcolor{BurntOrange}{\textbf{text}} pair under verification with external information (\textit{evidence}), both \textcolor{Orchid}{\textbf{images}} and \textcolor{LimeGreen}{\textbf{texts}}, collected from the web.
    The proposed framework retrieves and re-ranks the evidence while \methodname \ determines which pieces of information are most relevant to (support or refute) the image-text pair and then uses those to determine the pair's veracity.  }
    \label{figure:banner}
\end{figure*}

The dissemination of misinformation has greatly intensified in the digital age with the advent of the internet and social media platforms \cite{olan2022fake} and entails numerous adverse outcomes \cite{bennett2018disinformation, cantarella2023does, roozenbeek2020susceptibility, do2022infodemics, gamir2021multimodal, ruokolainen2020conceptualising}.
Generative AI advances, such as Large Language Models and Stable Diffusion, can now be exploited to create convincing and hyper-realistic misinformation \cite{zhou2023synthetic, xu2023combating}.
Nevertheless, decontextualization or image re-purposing, where images are taken out of their original context in order to support false narratives, remain a prevalent challenge in multimodal misinformation detection \cite{aneja2023cosmos, luo2021newsclippings}.
For instance, Fig.\ref{figure:banner} provides a real-world example, where a legitimate image is used as evidence to promote a false narrative, linking 5G technology to the transmission of COVID-19 when in reality, the image captures an anti-surveillance protest in Hong Kong
\footnote{\url{https://www.snopes.com/fact-check/5g-tower-torn-down-china-covid}}.

Recent years have witnessed a surge in research attention towards automated multimodal misinformation detection \cite{hangloo2022combating} with efforts focusing on the creation of multimodal misinformation datasets \cite{boididou2018verifying, nakamura2020fakeddit, aneja2023cosmos, luo2021newsclippings, jaiswal2017multimedia, papadopoulos2023synthetic, biamby2022twitter, muller2020multimodal} and detection methods \cite{khattar2019mvae, wang2018eann, yu2022bcmf, singhal2019spotfake, singhal2022leveraging, singhal2022leveraging}.
Nevertheless, solely relying on the analysis of an image-text pair is not always sufficient for detecting misinformation. More often that not, the incorporation and cross-examination of external information, i.e., evidence, is necessary \cite{guo2022survey}.
To this end, researchers have explored methods for automated and evidence-based fact-checking, initially focusing on text-based approaches \cite{wang2017liar, augenstein2019multifc, thorne2018fever, aly2021feverous} and, more recently, venturing into the realm of Multimodal Fact-Checking (MFC) \cite{yao2023end, suryavardan2023factify, abdelnabi2022open}.
Certain works on MFC, collect external evidence directly from fact-checked articles \cite{yao2023end, suryavardan2023factify}, which can be considered to be ``leaked evidence'' and introduce an unrealistic scenario: employing the work of human fact-checker as input while trying to ``early'' detect emerging new claims \cite{glockner2022missing}. 
Conversely, the NewsCLIPings+ dataset, comprises external information collected from the web \cite{abdelnabi2022open}, does not suffer from ``leaked evidence'' and has recently been used in numerous studies \cite{abdelnabi2022open, zhang2023ecenet, yuan2023support}.
Nevertheless, these works on MFC operated under the assumption that all collected external items from the web were relevant and only addressed the task of verdict prediction.

\textbf{Motivation.}
We aim to simulate a more realistic scenario, akin to what fact-checkers encounter daily, where multiple pieces of information are amassed and examined, but not all of them are necessarily relevant for supporting or refuting a claim. 
Thus, the challenge lies in effectively distinguishing between relevant and irrelevant evidence in order to assist the overall verdict prediction process.

\textbf{Contributions.}
We incorporate the step of Relevant Evidence Detection (RED), as part of the MFC process. 
Specifically, we propose the Relevant Evidence Detection Directed Transformer (\methodname) which comprises: ``Evidence re-ranking'', ``Modality Fusion'' and ``RED'' modules.
``Evidence re-ranking'' leverages intra-modal similarity to re-rank the evidence based on relevance and to create hard negative ``irrelevant'' samples.
``Modality Fusion'' leverages element-wise operations between the two modalities in order to cross-check their relation. 
For feature extraction and evidence re-ranking we experiment with CLIP ViT B/32 and L/14 \cite{radford2021learning} as well as ALBEF \cite{li2021align} and BLIP2 \cite{li2023blip}. 
Finally, ``RED'' examines all provided pieces of information, determines their relevance to the image-text pair and leverages the relevant ones to assess the pair's veracity.

We explore numerous variants of \methodname, with different architectures (e.g., single or dual stage) and mechanisms (e.g., guided attention). 
\methodname \ is optimized via multi-task learning on the NewsCLIPings+ dataset \cite{abdelnabi2022open} while its performance is also evaluated on the VERITE benchmark \cite{papadopoulos2023verite} comprising real-world multimodal misinformation. 
Finally, we propose the Out-Of-Distribution Cross-Validation (OOD-CV) evaluation protocol in order to capture models with superior generalizability across algorithmically created training data (NewsCLIPings+) and real-world data (VERITE). 

\textbf{Findings.}
We conduct extensive ablation and comparative experiments that demonstrate several key findings: (1) OOD-CV captures models with improved performance compared to conventional ``In-Distribution Validation'' (ID-V) (See Section \ref{sec:eval_prot}).
(2) By utilizing evidence re-ranking, a single piece of evidence per modality suffices while additional information can introduce noise and decrease overall performance. 
(3) The combination of element-wise operations can significantly boost detection accuracy, especially when contrasted with simple late-fusion concatenation.
(4) \methodname \ surpasses prior methods that do not incorporate external evidence by a substantial margin of up to 33.7\% and the \methodname-Baseline, which does not utilize the RED module, by 8.9\%.
Moreover, \methodname \ surpasses the current state-of-the-art on NewsCLIPings+ by up to 3\% without requiring multiple backbone encoders, numerous evidence or additional features. 

\section{Related Work}
\label{sec:related_work}

In recent years, automated misinformation detection has gained significant research attention, with ongoing exploration of methods for identifying false information across textual \cite{mridha2021comprehensive}, visual \cite{rana2022deepfake}, and multimodal formats \cite{alam2022survey}.
While several datasets have been developed for text-based fact-checking \cite{wang2017liar, augenstein2019multifc, thorne2018fever, aly2021feverous}, 
Glockner et al. \cite{glockner2022missing} showed that most employ ``leaked evidence'' from fact-checked articles. 
This leads to an unrealistic setting that is not applicable for early detection and emerging misinformation detection. 
In the context of multimodal misinformation, the majority of research only considers the image-text pairs \cite{jaiswal2017multimedia, sabir2018deep, wang2018eann, khattar2019mvae, singhal2019spotfake, yu2022bcmf, nakamura2020fakeddit, luo2021newsclippings, aneja2023cosmos, papadopoulos2023synthetic, papadopoulos2023verite, mu2023self} and do not incorporate external information or evidence; with few notable exceptions  \cite{yao2023end, suryavardan2023factify, abdelnabi2022open}.

MOCHEG was developed as an end-to-end dataset for MFC, including evidence retrieval, verdict prediction and explanation generation \cite{yao2023end}.
Top-k text evidence were retrieved using similarities from S-BERT and then re-ranked by a pre-trained BERT model while image evidence were retrieved based on cross-modal similarities from CLIP. Both S-BERT and CLIP were fine-tuned with contrastive learning.
However, while MOCHEG provides both text and image evidence, it only accommodates textual claims, not multimodal misinformation, so it can not be used for MFC. 
Moreover, the provided textual evidence can be considered ``leaked'' since they are collected solely from fact-check articles. 
In contrast, FACTIFY2 addresses multimodal evidence entailment as part of MFC, meaning the examination  of whether an article and an image entails the information to support or refute a claim \cite{suryavardan2023factify}. 
Recent studies relied on text summarization or top-k sentence retrieval from articles as part of the ``evidence retrieval'' process \cite{suryavardan2023findings}. 
However, FACTIFY2 only provides a single article and image as evidence and incorporates fact-checking articles for the ``Refute'' class, thus suffers from ``leaked evidence''. As a result, even simple baseline models can reach unrealistically high scores of 99-100\% on this class \cite{suryavardan2023findings}; limiting the practical usefulness of the dataset.
On the other hand, Abdelnabi et al.\cite{abdelnabi2022open} leverage the NewsCLIPings dataset \cite{luo2021newsclippings} and augment it with external information collected from Google API. We refer to this dataset as NewsCLIPings+ for simplicity. 
NewsCLIPings is a algorithmically created dataset, where image-text pairs are decontextualized using the multimodal encoder CLIP, as well as Scene and Person matching computer vision models. 
Since the ``falsified'' pairs in NewsCLIPings are algorithmically created and de-contextualized -individual modalities are legitimate but their combination is falsified- it secures the absence of ``leaked evidence''.

Recent works have utilized NewsCLIPings+ for MFC. 
The Consistency Checking Network (\textit{CCN}) \cite{abdelnabi2022open} leverages two attention-based memory networks for visual and textual reasoning, with ResNet152 and BERT, respectively, as well as a fine-tuned CLIP (ViT B/32) component for additional feature extraction.
The Stance Extraction Network (\textit{SEN}) leverages a similar architecture to CCN but with ResNet50 (pre-trained on Places365), ResNet152 (pre-trained on ImageNet) and S-BERT as the backbone encoders \cite{yuan2023support}. Additionally, it extracts the ``stance'' of external evidence towards the image-text pair and calculates the ``support-refutation score'' based on the co-occurrence of named entities between the claim and the textual evidence. 
Finally, the Explainable and Context-Enhanced Network (\textit{ECENet}) employs ResNet50, BERT and CLIP ViT-B/32 for feature extraction on images, texts, evidence and named entities within a ``Coarse- and Fine-grained Attention'' network for intra-modal and cross-modal feature reasoning \cite{zhang2023ecenet}. 
Nevertheless, while the aforementioned methods demonstrated promising results on NewsCLIPings+, they operated under the assumption that all collected information from the web are relevant, which is not necessarily valid.

One limitation of NewsCLIPings is that it relies on algorithmically created misinformation. 
This introduces uncertainty regarding the ability of models trained on it to generalize effectively to real-world multimodal misinformation.
For this reason, researchers have been experimenting with training multimodal misinformation detection models on algorithmically created datasets like NewsCLIPings and MEIR \cite{sabir2018deep} and evaluating their generalizability to real-world misinformation  \cite{papadopoulos2023synthetic}. 
Due to unimodal biases in widely adopted benchmarks, such as VMU Twitter \cite{boididou2018verifying} and COSMOS \cite{aneja2023cosmos}, we recently developed VERITE, an evaluation benchmark for real-world multimodal misinformation detection that accounts for unimodal biases \cite{papadopoulos2023verite}. However, VERITE and other similar evaluation datasets have not yet been used for MFC since they do not provide external evidence.

In light of the aforementioned considerations, we introduce a new architecture that incorporates ``Relevant Evidence Detection'' (RED) as a means of improving verdict prediction accuracy. 
We opt for training our model on NewsCLIPings+ while also evaluating its generalizability to VERITE, after augmenting it through evidence collection from the web.

\section{Methodology}
\label{sec:method}

\begin{table}
\centering
\scriptsize
\caption{Summary of important notations.}
  \begin{tabular}{|l|l|}

    \hline 
    \textbf{Term} & \textbf{Description} \\
    \hline
    $T^{v}_i$, $I^{v}_i$ & Text and image pair under verification \\
    $T^{e}_i$, $I^{e}_i$ & Text and image evidence related to the pair \\        
    $\mathcal{T}^+_i$, $\mathcal{I}^+_i$ & Set of relevant text and image evidence related to the pair \\ 
    $\mathcal{T}^-_i$, $\mathcal{I}^-_i$ & Set of irrelevant text and image evidence related to the pair \\     

    $\mathcal{T}^v$, $\mathcal{I}^v$ & Collection of all texts and images under verification \\
    $\mathcal{T}^e$, $\mathcal{I}^e$ & Collection of all text and image evidence \\
    
     $\mathcal{E}^+$, $\mathcal{E}^-$ & All relevant and irrelevant evidence after re-ranking\\
     $\mathcal{E}$ & Combined, ranked and permuted relevant and irrelevant evidence\\
     $M, K$ &  Number of text and image evidence\\
     $y^v$, $\boldsymbol{y}^e$ & Labels for verdict and evidence relevance \\

     $F_{T^v}$, $F_{I^v}$ & Extracted features for the texts and images under verification \\
     $F^v$ &  Fused multimodal features for the pairs under verification \\
     $F_{\mathcal{E}}$ & Extracted features for the external evidence \\
     $dim$ & Dimensionality of the backbone encoder \\
     $D(\cdot)$ & Transformer encoder \\
     $d$ & The output of the Transformer \\
     $a$ & Attention scores \\
     $CLS$ & Classification token \\

     
    \hline

\end{tabular}
\label{table:notation}    
\end{table}

\subsection{Problem Formulation}

We define the tasks of Multimodal Fact-Checking (MFC) and Relevant Evidence Detection (RED) as follows: given a dataset  ${(I^{v}_i, T^{v}_i, \mathcal{I}^{e}_i}, \mathcal{T}^{e}_i, y^{v}_i, \boldsymbol{y}^{e}_i)_{i=1}^{N}$,
where 
$(I^{v}_i, T^{v}_i)$
represents the image-text pair under verification (pair, from now on), 
$\mathcal{I}^{e}_i = [I^{e}_{i1}, I^{e}_{i2}, \ldots, I^{e}_{i2K}]$
and 
$\mathcal{T}^{e}_i = [T^{e}_{i1}, T^{e}_{i2}, \ldots, T^{e}_{i2M}]$
represent an array of image evidence with $2\cdot K$ elements ($K$ relevant + $K$ irrelevant) and text evidence with $2\cdot M$ elements ($M$ relevant + $M$ irrelevant), respectively,
$y^{v}_i \in {0, 1}$ denotes the overall verdict label, truthful (0) or misinformation (1) and 
$\boldsymbol{y}^{e}_i = (y^{e}_{i1}, y^{e}_{i2}, \ldots, y^{e}_{i2\cdot(M+K)})$ is an array of binary labels of $2\cdot (M+K)$ size denoting whether each piece of evidence is relevant (1) or irrelevant (0) to the pair;
with the primary objective of training a classifier $f: (\mathcal{I}^v, \mathcal{T}^v, \mathcal{I}^e, \mathcal{T}^e) \rightarrow (\hat{y}^v, \hat{\boldsymbol{y}}^e)$.
Table \ref{table:notation} summarizes key notations of this paper.

\subsection{Evidence Retrieval and Re-ranking}
\label{sec:ev_retrieval}

To create a relevant evidence training set, we leverage the NewsCLIPings+ dataset \cite{abdelnabi2022open} comprising 85,360 image-text pairs, balanced in terms of truthful and misinformation pairs.
The authors use the text $T^v_i$ to retrieve visual evidence $\mathcal{I}^e_i$ and the image $I^v_i$ to retrieve textual evidence $\mathcal{T}^e_i$. 
Each pair is associated with up to 19 textual evidence and up to 10 visual evidence, with a total of 146,032 text evidence and 736,731 image evidence; collected via Google API.
These items are not ``robust evidence'' in the sense that they necessarily support or refute the claim but rather ``potential evidence'', nevertheless, we use the term “evidence” for consistency with most papers in the field.
While some level of relevance between the collected evidence and the pair expected, we undertake the task of re-ranking all gathered evidence to heighten the likelihood of their relevance.

Following \cite{abdelnabi2022open}, we use the pre-trained CLIP ViT B/32 \cite{radford2021learning} 
as the backbone encoder in order to extract visual $F_I\in{R}^{dim\times 1}$ and textual features $F_T\in{R}^{dim\times 1}$ with dimensionality $dim=512$, for $\mathcal{I}^{v}, \mathcal{I}^{e}$ and $\mathcal{T}^{v}, \mathcal{T}^{e}$.
Additionally, we experiment with ALBEF Base \cite{li2021align}, BLIP2 ViT L \cite{li2023blip} and CLIP ViT L/14  with $dim=768$, all taken from the LAVIS\footnote{\url{https://github.com/salesforce/LAVIS}} library. 
We signify the ranked or ``relevant'' evidence with the superscript ``+'', and rank them based on the intra-modal cosine similarity $sim$ as follows: 
\begin{equation}
\label{eqn:relevant_rank}
\begin{split}
 \mathcal{I}^+_i = \operatorname*{argsort}\limits_{I^e_j\in \mathcal{I}^e_{i}}sim(F_{I^{v}_i}, F_{I^{e}_j}) \\
 \mathcal{T}^+_i = \operatorname*{argsort}\limits_{T^e_j\in \mathcal{T}^e_{i}}sim(F_{T^{v}_i}, F_{T^{e}_j}) 
\end{split}
\end{equation}

\begin{figure}[t]
  \centering
   \includegraphics[width=0.7\linewidth]{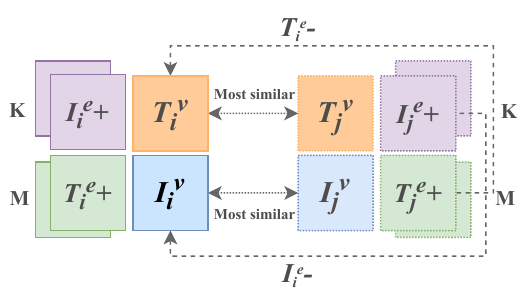}

   \caption{Visualization of Eq.\ref{eqn:irrelevant_rank}. Hard negative sampling for retrieving ``irrelevant'' evidence.}
   \label{fig:evidence_retrieval}
\end{figure}

For the ``irrelevant'' evidence class, denoted by the superscript ``-'', we employ hard negative sampling, instead of depending on random sampling.
More specifically, 
we calculate the most similar item based on text-text similarity, fetch its ranked evidence and employ them as $\mathcal{I}^-_i$, and use image-image similarity for $\mathcal{T}^-_i$. 
We incorporate the `opposite' modality in similarity calculations in order to mimic the evidence retrieval process of NewsCLIPings+. 
The process of hard negative sampling ``irrelevant'' evidence is illustrated in Fig.\ref{fig:evidence_retrieval} and can be expressed as follows:
\begin{equation}
\label{eqn:irrelevant_rank}
\begin{split}
\mathcal{I}^-_i = \mathcal{I}^+_j,\text{ where }j = \operatorname*{argmax}\limits_{T^v_j\in \mathcal{T}^v, j \neq i}sim(F_{T^{v}_i}, F_{T^{v}_j})
 \\
 \mathcal{T}^-_i = \mathcal{T}^+_j,\text{ where }j = \operatorname*{argmax}\limits_{I^v_j\in \mathcal{I}^v, j \neq i}sim(F_{I^{v}_i}, F_{I^{v}_j})
\end{split}
\end{equation}
\noindent We leverage efficient indexing (Meta's FAISS\footnote{\url{https://ai.meta.com/tools/faiss}}) to improve scalability and speed. 

After retrieving relevant evidence $\mathcal{E}^+ = (\mathcal{T}^+, \mathcal{I}^+)$ 
and irrelevant $\mathcal{E}^- = (\mathcal{T}^-, \mathcal{I}^-)$, we also define $\boldsymbol{y}^e+ = [1, 1, \ldots,1]$ and $\boldsymbol{y}^e- = [0, 0, \ldots,0]$, each having size M+K. 
In order to avoid overfitting based on positional patterns during training, we combine all evidence 
$\mathcal{E'} = \mathcal{E}^+ \cup \mathcal{E}^-$
and labels 
$\boldsymbol{y'}^e = [1, 1, ... 1, 0, 0, ..., 0]$, shuffle their positions with $P$ representing the permutation positions of the elements and the final evidence and labels become: 
$\mathcal{E} = \mathcal{E'}[P]$ and 
$\boldsymbol{y}^e = \boldsymbol{y'}^e[P]$.
Evidence $\mathcal{E}$ should consist of $2 \cdot (M+K)$ items.
For instance, setting $M=1$ signifies our expectation of having 1 relevant and 1 irrelevant textual evidence.
If this requirement is not met, we randomly sample additional items from the dataset.

\subsection{Modality Fusion module}

To enhance the cross-examination of image-text pairs and accentuate specific consistencies or inconsistencies, we utilize a range of multimodal fusion operations. 
In addition to simple concatenation of the two modalities' features, denoted as $[F_{I^v};F_{T^v}]$, we also employ ``addition'', ``subtraction'' and ``multiplication''.
Our rationale is that ``addition'' emphasizes complementarity, ``subtraction'' accentuates differences and ``multiplication'' underscores shared aspects.
We express the modality fusion module as follows:
\begin{equation}
\label{eqn:fusion}
F^v = [F_{I^v};F_{T^v};F_{I^v}+F_{T^v};F_{I^v}-F_{T^v};F_{I^v}*F_{T^v}]
\end{equation}
with $F^v\in{R}^{dim\times5}$. 
Prior research in multimodal fusion has explored different element-wise operations, including multiplication \cite{koutlis2023memefier} and outer product \cite{kumar2022hate}.
However, to the best of our knowledge, the combination of multiple fusion operations remains unexplored, especially for MFC.

\begin{figure*} 
    \centering
  \subfloat[\label{fig:transformer}]{%
       \includegraphics[width=0.70\linewidth]{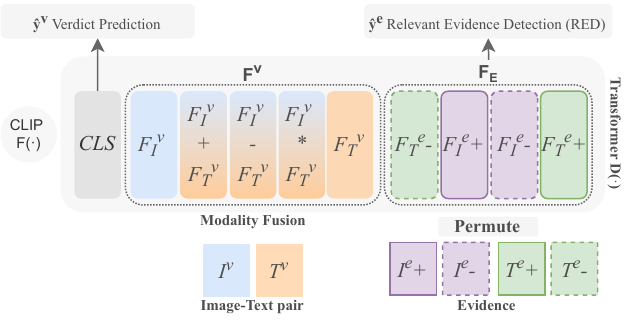}}
    \hfill
  \subfloat[\label{fig:ssl_vs_dsl}]{%
        \includegraphics[width=0.30\linewidth]{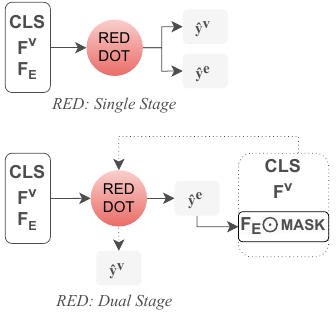}}
    \\
  \caption{(a) Overview of the proposed Transformer $D(\cdot)$ in \methodname, employing ``Modality Fusion''. (b)
  High-level overview of the single and dual stage \methodname \ variants. Dotted lines represent the second stage in DSL.
  }
  \label{fig:method} 
\end{figure*}

\subsection{Verdict Prediction module}

We employ a Transformer encoder $D(\cdot)$ to obtain the predicted verdict $\hat{y}^{v}$. Our \methodname-Baseline experiments that only leverage relevant evidence are expressed by Eq.\ref{eqn:d_cls_fv_e} while the final verdict is predicted by Eq.\ref{eqn:verdict_classifier}.
\begin{equation}
\label{eqn:d_cls_fv_e}
[\mathbf{d_{\text{CLS}}, d_{F^v}, d_{F_{\mathcal{E+}}}}] = D([\text{CLS}; F^v;F_{\mathcal{E+}}])
\end{equation}
\begin{equation}
\label{eqn:verdict_classifier}
\hat{y}^{v}=\textbf{W}_1\cdot \text{GELU}(\textbf{W}_0\cdot \text{LN}(\mathbf{d}_{\text{CLS}}))
\end{equation}
where LN stands for Layer Normalization, CLS stands for classification token, $\textbf{W}_0\in\mathbb{R}^{1 \times dim}$ 
is a GELU activated fully connected layer and 
$\textbf{W}_1\in\mathbb{R}^{\frac{dim}{2} \times 1}$ is the final verdict classification layer.
$D(\cdot)$ is optimized for binary classification based on the binary cross entropy (BCE) loss function $L^v$, after applying the sigmoid activation function on $\hat{y}^v$.

\subsection{Relevant Evidence Detection}

The Relevant Evidence Detection (RED) module is responsible for discerning which evidence in $\mathcal{E}$ are relevant or irrelevant to the claim under verification.
Given that our work introduces and explores the RED task for the first time, we deem important to investigate diverse architectural approaches which can also serve as valuable ``strong baselines'' for future research in this field.
All methods utilize Transformer $D(\cdot)$ as shown in Eq.\ref{eqn:D_with_E} with $\mathbf{d}_{F_{\mathcal{E}}}$ as shown in Eq.\ref{eqn:d_E} and Eq.\ref{eqn:verdict_classifier} to predict $\hat{y_v}$. 
Moreover, all methods leverage multi-task learning to predict both binary $\hat{y}^v$ and multi-label $\boldsymbol{\hat{y}}^e$ and are optimized based on $L = L^v + L^e$ with $L^e$ being the average BCE loss for multi-label classification, after applying the sigmoid activation function on $\boldsymbol{\hat{y}}^e$.
An overview of the \methodname \ architecture can be seen in Fig.\ref{fig:method}.
\begin{equation}
\label{eqn:D_with_E}
[\mathbf{d_{\text{CLS}}, d_{F^v}, d_{F_{\mathcal{E}}}}] = D([CLS;F^v;F_{\mathcal{E}}]) 
\end{equation}
\begin{equation}
\label{eqn:d_E}
\mathbf{d}_{F_{\mathcal{E}}} = [\mathbf{d_{F_{\mathcal{E}_1}}}, 
\dots, 
\mathbf{d_{F_{\mathcal{E}_{(M+K)\cdot2}}}}]
\end{equation}

\textbf{(1) Single-stage Learning (SSL)}: employs Eq.\ref{eqn:D_with_E} and Eq.\ref{eqn:y_e_classifier} to predict $\boldsymbol{\hat{y}}^e$ in a single stage. 
\begin{equation}
\label{eqn:y_e_classifier}
\footnotesize
\hat{y}^{e} = [\textbf{W}_3 \cdot \text{GELU}(\textbf{W}2 \cdot \text{LN}(\mathbf{d_{F_{\mathcal{E}_i}}}))], \quad \text{for } i = 1 \text{ to } 2 \cdot (M+K)
\end{equation}
with $\textbf{W}_2\in\mathbb{R}^{1 \times dim}$ 
is a GELU activated fully connected layer and 
$\textbf{W}_3\in\mathbb{R}^{256 \times 1}$.

\textbf{(2) Single-stage Learning with Guided Attention (SSL+GA)}: Similar to SSL but instead of Eq.\ref{eqn:y_e_classifier}, we apply a type of ``guided attention'' where the $L_v$ is directly applied onto the attention weights.
Consider the vector $\mathbf{d} = [\mathbf{d_{\text{CLS}}, d_{F^v}, d_{F_{\mathcal{E}}}}]$, to calculate the attention scores, we use Eq.\ref{eqn:attention_scores} and Eq.\ref{eqn:y_e_guided} to predict $\boldsymbol{\hat{y}}^e$
\begin{equation}
\label{eqn:attention_scores}
\mathbf{a} = \frac{\mathbf{d} \cdot \mathbf{d^T}}{dim}
\end{equation}
\begin{equation}
\label{eqn:y_e_guided}
\hat{y}^{e} = \mathbf{a}[<CLS>][-2\cdot(M+K):]
\end{equation}
where $<CLS>$ represents the position of the classification token, providing a global representation of attention and $-2\cdot(M+K)$ denotes the last $2\cdot(M+K)$ items in $\textbf{d}$, corresponding to the evidence.

\textbf{(3) Dual-stage Learning (DSL)}:
first employs Eq.\ref{eqn:D_with_E} and Eq. \ref{eqn:y_e_classifier}. Afterwards, evidence $\mathcal{E}$ are masked by $\text{MASK} \in \{0, 1\}^{dim \times 2\cdot(M+K)}$ where 0s denote predicted irrelevant and 1s predicted relevant evidence. 
During training, we employ teacher enforcing. 
At the second stage, we apply $\text{MASK}$ onto the evidence and re-process with Transformer $D(\cdot)$ with Eq.\ref{eqn:D_with_E_mask} and then use Eq.\ref{eqn:verdict_classifier} to predict $\hat{y}^v$.
\begin{equation}
\label{eqn:D_with_E_mask}
[\mathbf{d_{\text{CLS}}, d_{F^v}, d_{F_{\mathcal{E}}}}] = D([\text{CLS}; F^v;F_{\mathcal{E}} \odot \text{MASK}])
\end{equation}

\textbf{(4) Dual-stage Learning with Guided Attention (DSL+GA)}: similar to DSL, but in the first stage it utilizes guided attention (as in Eq.\ref{eqn:attention_scores} and Eq.\ref{eqn:y_e_guided}) instead of Eq.\ref{eqn:y_e_classifier} to predict $\boldsymbol{\hat{y}}^e$.
The second stage remains identical to DSL. 

\textbf{(5) Dual-stage Learning with two Transformers (DSL+D2)}: similar to DSL, but in the second stage we employ a second identical Transformer encoder $D2(\cdot)$ with $[\mathbf{d_{\text{CLS}}, d_{F^v}, d_{F_{\mathcal{E}}}}] = D2([\text{CLS}; F^v;F_{\mathcal{E}} \odot \text{MASK}])$. 
The first stage remains identical to DSL. 

\section{Experimental Setup}
\label{sec:experimental}

\subsection{Evaluation Protocol}
\label{sec:eval_prot}

Our goal is to examine the generalizability of \methodname \ to real-world misinformation. For this reason, we train \methodname \ on the NewsCLIPings+ dataset and evaluate its performance on the Out-Of-Context (OOC) pairs from the VERITE benchmark. 
To gather ``external evidence'' via Google API for VERITE, we followed the process outlined in \cite{abdelnabi2022open}; which we make publicly available in order to facilitate future research and ensure fair comparability across studies. 

Our evaluation protocol departs from ``In Distribution Validation'' (ID-V) which involves validation and check-pointing on NewsCLIPings and final testing on VERITE \cite{papadopoulos2023verite}.
Instead, we leverage ``Out-Of-Distribution Cross Validation'' (OOD-CV) with k-fold cross-validation and check-pointing directly on a VERITE fold (k-fold=3), while training \methodname\ on the NewsCLIPings+ training set.
Partially inspired by \cite{koh2021wilds}, we hypothesize that due to NewsCLIPings+ and VERITE following different distributions -the first comprising algorithmically created samples and the latter real-world misinformation- OOD-CV can capture a version of the model with improved generalizability onto the ``out-of-distribution'' VERITE.

We report ``Accuracy'' on NewsCLIPings+ based on ID-V and ``True vs. OOC Accuracy'' and the standard deviation on VERITE based on OOD-CV.
During evaluation of \methodname \ on VERITE we solely employ ``M,K+'' items retrieved by Google API and re-ranked by our pipeline, without injecting ``negative evidence''. The model has to determine which items are relevant, simulating automated fact-checking ``in the wild''. 

\subsection{Competing Methods}
\label{sec:competing_methods}

On NewsCLIPings+, we compare \methodname \ against \textit{\textbf{CCN}} \cite{abdelnabi2022open}, \textit{\textbf{ECENet}} \cite{zhang2023ecenet} and \textit{\textbf{SEN}} \cite{yuan2023support} which incorporate external evidence. For more information about these models, see Section \ref{sec:related_work}. 
Moreover, we include methods that do not incorporate external evidence, namely: CLIP \cite{longoni2022news} and Self-Supervised Distilled Learning (\textit{\textbf{SSDL}}) \cite{mu2023self} and the Detector Transformer (\textit{\textbf{DT}}) \cite{papadopoulos2023synthetic}.
On VERITE, we compare \methodname \ against the Detector Transformer (\textbf{\textit{DT}}) \cite{papadopoulos2023verite} employing features from CLIP ViT-L/14 after being trained on ``R-NESt + CHASMA-D + NC-t2t''; a combination of three different algorithmically created datasets with 437,673 samples in total. 
To ensure comparability, we faithfully replicate the \textit{DT} and train it on NewsCLIPings+ without external evidence. 
Within our framework, the \textit{DT} can be expressed as $D([\text{CLS}; F_{I^v}; F_{T^v}])$ with features from CLIP ViT-L/14 and is trained under the ID-V protocol.  

\subsection{Implementation Details}

We train \methodname\ for a maximum of 100 epochs with early stopping at 10 epochs, using the Adam optimizer and a learning rate of $lr = 1e-4$.
When tuning hyperparameters, we consider the following configurations: $t \in \{4, 6\}$ transformer layers, $z\in\{128, 2048\}$ for the dimension of the feed-forward network, and $h\in\{2, 8\}$ for the number of attention heads. 
The dropout rate is set at 0.1 and the batch size to 512. 
For faithful reproduction of \textit{DT}, we follow the implementation details of \cite{papadopoulos2023verite} which uses $lr = 5e-5$ and $t \in \{1,4\}$. 
Finally, to ensure experiment reproducibility, we use a constant random seed (0) for PyTorch, Python random, and NumPy.

\section{Experimental Results}
\label{sec:results}

\subsection{Out-of-distribution Cross-Validation} 
Table \ref{table:m_k} showcases the performance of \methodname \ variants, with $M=1$ texts and $K=1$ images as evidence, when trained and validated under two different protocols: ID-V or the proposed OOD-CV. 
We observe that OOD-CV consistently gives better results than ID-V, with 72.3\% and 70.4\% mean accuracy, respectively. 
These results suggest that OOD-CV is more effective at capturing a model with better generalizability from algorithmically created samples (NewsCLIPings+) to real-world out-of-context misinformation (VERITE).
Based on these findings, we recommend that future research endeavors involving VERITE adopt the OOD-CV protocol to ensure a more robust and representative model performance on the dataset. We employ OOD-CV in all following experiments. 

\begin{table}[!t]
\centering
\caption{
Variants of \methodname\ (w/ CLIP ViT B/32) trained under the OOD-CV protocol with M texts and K images as evidence.
We report Accuracy on VERITE and the standard deviation (in parentheses) for OOD-CV (k-fold=3). 
For $M,K=1$ we also report detection Accuracy under ID-V.  
\underline{Underline} denotes the best performance per method.  
}
  \begin{tabular}{|l|l|l|l|l|}

    \hline 
    \textbf{Method} &
    \textbf{M,K=1 (ID-V)} &    
    \textbf{M,K=1} &
    \textbf{M,K=2} &
    \textbf{M,K=4}   
    \\

    \hline

    Baseline & 
    69.7 &
    \underline{70.7} (1.5) &
    65.9 (2.3) & 
    66.4 (2.3) \\

    SSL & 
    70.0 &
    71.8 (0.1) &
    \underline{72.1} (3.7) & 
    67.2 (1.8) \\

    SSL + GA & 
    71.2 &
    \underline{72.6} (1.7) &
    71.9 (3.0) &
    65.8 (2.7) \\

    DSL + D2 & 
    69.1 &
    \underline{71.3} (0.5) &
    66.5 (2.6) &
    65.3 (0.7) \\

    DSL + GA & 
    71.3 &
    \underline{73.6} (1.7) &
    69.9 (2.5) &
    62.7 (1.7) \\

    DSL & 
    70.9 &
    \underline{73.9} (0.5) &
    70.0 (2.4) &
    64.7 (1.6) \\

    \hline

    \textcolor{gray}{Mean} &
    \textcolor{gray}{70.4} &    
    \textbf{72.3} \textcolor{gray}{(1.0)} &
    \textcolor{gray}{69.4 (2.8)} &
    \textcolor{gray}{65.3 (1.8)} \\
    
    \hline
\end{tabular}
\label{table:m_k}
\end{table}

\begin{table}
\centering
\caption{
Ablation of \methodname\ (w/ CLIP ViT B/32) 
with different Modality Fusion combinations. 
$F^v$ followed by `-' signifies removal of one fusion operation.
We report Accuracy on VERITE and the standard deviation (in parentheses). 
}
  \begin{tabular}{|l|l|l|l|l|l|}

    \hline 
    \textbf{} &
    \textbf{$F_{I^v};F_{T^v}$} &
    \textbf{$F^v$} &    
    \textbf{$F^v$-``-''} &
    \textbf{$F^v$-``+''} &
    \textbf{$F^v$-``*''}
    \\

    \hline

    Baseline
    & 
    66.7 (0.9)	&
    70.7 (1.5)	&
    68.3 (1.2)	&
    70.3 (1.7)	&
    \underline{71.6} (4.1) \\

    SSL & 
    71.6 (1.2) &
    71.8 (0.1) &
    73.2 (1.2)	&
    \underline{73.9} (1.1) &
    72.9 (0.3) \\

    SSL+GA &
    70.6 (0.8) & 
    \underline{72.6} (1.7) &
    72.0 (1.5) &
    71.6 (2.4) &
    70.5 (2.1) \\

    DSL+D2 & 
    68.8 (1.6) &
    \underline{71.3} (0.5) &
    70.0 (0.8) &
    69.8 (1.7) &
    69.3 (1.7) \\

    DSL+GA &
    69.8 (1.1) &
    \underline{73.6} (1.7) &
    71.2 (2.0) &
    \underline{73.6} (1.7) &
    72.8 (1.1) \\

    DSL &
    71.0 (0.4) &
    73.9 (0.5) &
    71.5 (2.2) &
    72.9 (1.1) &
    \underline{74.2} (1.5) \\

    \hline
    
    \textcolor{gray}{Mean} & 
    \textcolor{gray}{69.8 (1.0)} &
    \textbf{72.3} \textcolor{gray}{(1.0)} &
    \textcolor{gray}{71.0 (1.5)} &
    \textcolor{gray}{72.0 (1.6)} &
    \textcolor{gray}{71.8 (1.8)} \\
    
    \hline

\end{tabular}
\label{table:fusion_ablation}    

\end{table}

\subsection{Impact of multiple ranked evidence}
Table \ref{table:m_k} also illustrates the impact of leveraging $M, K \in \{1, 2, 4\}$ ranked texts and images as evidence. 
We observe that the best performance is yielded with $M,K=1$, with an average accuracy of 72.3\%, while the performance tends to decline with the inclusion of additional evidence. 
\methodname \ is explicitly trained to detect evidence relevance. 
Setting ``M,K+ $>$ 1'' introduces additional but less relevant items in $\mathcal{E+}$ while their labels are defined as relevant ($\boldsymbol{y}^e+$), thus introducing noise that tends to perplex the network and degrade its performance. 
Similarly, we observe that even \methodname-Baseline that does not leverage RED and, consequently, irrelevant evidence, still demonstrates a reduction in accuracy with the addition of more items. 
We can infer that the process of re-ranking evidence, as described in Section \ref{sec:ev_retrieval}, can be advantageous compared to leveraging all collected evidence both in terms of detection accuracy and decreased computational complexity; since the network has fewer items to process. 
We employ $M,K=1$ in all following experiments.

\subsection{Ablation on Modality Fusion}
Table \ref{table:fusion_ablation} presents the ablation results for different Modality Fusion configurations within \methodname \ variants. 
Our findings indicate that the simple concatenation ($F_{I^v};F_{T^v}$) results in the lowest average performance (69.8\%) across all \methodname \ variants while employing all fusion operations ($F^v$) produces the highest (72.3\%).
However, there are individual instances where alternative fusion operations demonstrate superior performance. 
For instance, \methodname-DSL \ attains the highest overall accuracy while utilizing $F^v-\text{``*''}$ (removing the multiplication operation), yielding 74.2\% and surpassing the 73.9\% accuracy achieved with $F^v$.
Notably, \methodname-SSL yields 73.9\% without ``addition'' ($F^v-\text{``+''}$) and 71.8\% with $F^v$; a +2.9\% improvement. 
Based on these outcomes, we can conclude that leveraging $F^v$ can lead to adequate results, but if optimal performance is required, it is advisable to experiment with multiple multimodal fusion configurations.

\begin{table*}
\centering
\scriptsize
\caption{
Comparison with the state-of-the-art.
All methods are trained on NewsCLIPings with the exception of $\dagger$ denoting ``R-NESt + CHASMA-D + NC-t2t'' \cite{papadopoulos2023verite}.
Here, $M,K+$ and $M,K-$ represent the number of relevant and irrelevant evidence, respectively.
We report Accuracy on NewsCLIPngs+ and ``True vs OOC'' Accuracy as well as the standard deviation (in parentheses) on VERITE. 
}
  \begin{tabular}{|l|l|l|l|l|l|}
    \hline 
    \textbf{Method} &
    \textbf{Encoder} &
    \textbf{M,K+} & 
    \textbf{M,K-} & 
    \textbf{NewsCLIPings} 
    & \textbf{VERITE} 
    \\    
    \hline
    \textit{CLIP} \cite{abdelnabi2022open} & CLIP ViT B/32 (fine-tuned) & 0 & 0 & 60.2 & - \\
    
    DT \cite{papadopoulos2023synthetic} & CLIP ViT L/14 & 0 & 0 & 65.7 & - \\

    SSDL \cite{mu2023self} & 
    CLIP ViT B/32
    & 0 & 0 & 67.0 & - \\
    
    SSDL \cite{mu2023self} & 
    CLIP RN50 (fine-tuned) &
    0 & 0 & 71.0 & - \\

    \textit{DT}  \cite{papadopoulos2023verite} & CLIP ViT L/14 &
    0 & 0 & - & 57.5 \\ 

    \textit{DT} \cite{papadopoulos2023verite} $\dagger$
    & CLIP ViT L/14 &
    0 & 0 & - & 72.7 \\    

    \methodname-Baseline & ALBEF Base & 0 & 0 & 54.6  & 54.9 (3.4) \\
    \methodname-Baseline & BLIP2 ViT L &  0 & 0 & 68.2  & 61.1 (2.4) \\
    \methodname-Baseline & CLIP ViT B/32 &  0 & 0 & 71.6 & 63.7	(2.8) \\    
    \methodname-Baseline & CLIP ViT L/14 & 0 & 0 & 81.7 & 75.5 (3.7) \\
    \hline    

    \textit{CCN}	\cite{abdelnabi2022open} & CLIP ViT B/32 (fine-tuned) + ResNet152 + BERT (fine-tuned).
    & 
     ALL & 0 & 84.7 & - \\

     \textit{SEN} \cite{yuan2023support} & ResNet50 (Places365), ResNet152 (ImageNet), S-BERT
     & ALL & 0 & 87.1 & - \\

    \textit{ECENet} \cite{zhang2023ecenet} & 
    CLIP ViT B/32. BERT, ResNet50
    & ALL & 0 & 87.7 & - \\

    \hline
    
    \methodname-Baseline & ALBEF Base & 1 & 0 & 80.8 & 71.5	(0.3) \\
    
    \methodname-Baseline & BLIP2 ViT L&  1 & 0 & 85.7	& 71.5 (3.0) \\

    \methodname-Baseline & CLIP ViT B/32 &
    1 & 0 & 87.8 & 70.7 (1.5) \\

    \methodname-Baseline & CLIP ViT L/14 &
    1 & 0 & \textbf{90.3 }& 70.6 (3.1) \\
    
    \hline
    \methodname-DSL & ALBEF Base & 1 & 1 & 80.1	& 72.0	(2.3) \\
    \methodname-DSL & BLIP2 ViT L &  1 & 1 & 84.8 & 72.7	(2.6) \\

    \methodname-DSL & CLIP ViT B/32 &
    1 & 1 & 84.5 & 
    73.9 (0.5)
    \\

    \methodname-SSL & 
    CLIP ViT L/14 & 
    1 & 1 
    & 87.9	& 
    \textbf{76.9} (5.4)
    \\   

    \hline
    
\end{tabular}
\label{table:sota}
\end{table*}

\begin{table}
\centering
\caption{Point-biserial correlation coefficient (r) and p-values (p) (in parentheses) applied on the similarity between various features extracted either with CLIP B/32 or L/14. 
\underline{Underlines} denote relations without statistical significance ($p>0.05$). 
}
  \begin{tabular}{|l|l|l|l|l|}

    \hline 
    \textbf{Similarity} &
    \multicolumn{2}{|l|}{\textbf{NewsCLIPings+}} &
    \multicolumn{2}{|l|}{\textbf{VERITE}}
    \\

    \hline

    & B/32 & L/14 & B/32 & L/14 \\

    \hline

    $F_{I^v}$, $F_{T^v}$  & -0.38 (0.0) & -0.64 (0.0) & -0.46 (1.4e-35) & -0.59 (6.2e-63) \\

    \hline
    $F_{I^v}$, $F_{I^e}$ 
    & -0.51 (0.0) 
    & -0.55 (0.0) & -0.30 (0.1e-15) 
    & -0.33 (2.2e-18)
    \\
    
    \hline
    $F_{T^v}$, $F_{T^e}$ 
    & -0.28 (0.0) 
    & -0.37 (0.0)
    & \underline{-0.03 (0.47)} 
    & \underline{-0.03 (0.42)} \\
    
    \hline
    $F_{T^e}$, $F_{I^e}$ 
    & -0.17 (0.0)
    & -0.28 (0.0)
    & \underline{-0.03 (0.48)}
    & \underline{-0.07 (0.08)}
    \\
 
    \hline

\end{tabular}
\label{table:correlation}    
\end{table}

\subsection{Comparative Study: NewsCLIPings}
Table \ref{table:sota} presents the comparison between \methodname \ and the current state-of-the-art on NewsCLIPings+ and VERITE.
In regards to NewsCLIPings without external evidence, we observe that \methodname-Baseline (M,K+=0) outperforms prior methods, namely CLIP (60.2), DT (65.7) and SSDL (67.1), scoring 71.6\% and 81.7\% while using CLIP ViT B/32 and L/14 respectively. 
All other factors being equal, this can be primarily attributed to our element-wise modality fusion integrated in the Transformer-based architecture. 
Moreover, we observe that all methods that leverage external information significantly outperform methods that do not. 
With CLIP ViT B/32 as the backbone encoder, \methodname-Baseline (M,K+=1) yields 87.8\% on NewsCLIPings+, competing with and even outperforming the SotA, namely CCN (84.7\%), SEN (87.1\%) and ECENet (87.7\%), while only employing a single piece of evidence per modality and without requiring fine-tuning of the backbone encoder, nor multiple encoders (e.g., BERT and ResNet on top of CLIP) nor additional features (e.g., named entities).
Similarlty, with CLIP ViT L/14, \methodname-Baseline (M,K+=1) significantly outperforms the SotA, reaching 90.3\% Accuracy, translating into a +3.0\% relative improvement over ECENet. 
This is followed by \methodname-SSL, reaching 87.9\% on NewsCLIPings+ and striking the best balance between high performance on both NewsCLIPings+ and VERITE. 
These results highlight the advantage of the proposed ``Evidence re-ranking'' and ``Modality Fusion'' modules.

\subsection{Comparative Study: VERITE}
Similar to NewsCLIPings, we observe that \methodname \ leveraging CLIP ViT L/14 as the backbone yields consistently higher performance than the alternatives on VERITE.  
More specifically, \textit{Detector Transformer (DT)} (M,K+=0) trained on NewsCLIPings+ with features from CLIP ViT L/14 yields 57.5\% accuracy on VERITE while \methodname-Baseline (M,K+=0) reaches 75.5\% without external evidence surpassing its counterparts that leverage ALBEF (54.6), BLIP2 (68.2) or CLIP ViT B/32 (63.7). 
We can deduce that the selection of the backbone encoder plays a pivotal role. 
However, the most substantial difference between \textit{DT} and \methodname-Baseline is that the latter leverages our proposed modality fusion module, which demonstrates a noteworthy enhancement in performance.
On top of that, by leveraging CLIP ViT L/14 and the proposed Relevant Evidence Detection module, \methodname-SSL can further improve performance up to 76.9\%, a notable +33.7\% relative improvement over \textit{DT}, +1.9\% over \methodname-Baseline without evidence and +8.9\% with evidence. 
With CLIP B/32, \methodname-DSL outperforms \textit{DT} by a notable +28.5\%,  \methodname-Baseline without evidence by +16\% and with evidence by +4.5\%. 

We observe that integrating external evidence into \methodname-Baseline tends to improve performance on VERITE while employing 3 out of 4 encoders, namely ALBEF, BLIP2, CLIP ViT B/32. 
However this is not the case with CLIP ViT L/14, with which the performance of \methodname-Baseline decreases on VERITE despite achieving the highest overall performance on NewsCLIPings+. 
Since CLIP is a ``black box'' neural network we can not definitively explain this result. Nevertheless, in order to interpret this behavior, we calculate the correlation matrices on the similarity scores between various features, with either CLIP B/32 or L/14, as seen in Table \ref{table:correlation}. 
We observe that somewhat different patterns govern NewCLIPings and VERITE. 
While the correlation between image-text ($F_{I^v}, F_{T^v}$) similarities and the target variable increases with L/14 on both datasets, the relations between texts and textual evidence ($F_{T^v}, F_{T^e}$) as well as visual-evidence and textual-evidence  ($F_{I^e}, F_{T^e}$) are tenuous in VERITE and relatively strong in NewsCLIPings+.
Therefore, when \methodname-Baseline is trained on NewsCLIPings with L/14 and without evidence, it relies on the strong signal provided by ``image-text'' pairs and achieves high performance on both datasets. 
Thereafter, integrating external information helps improve performance on NewsCLIPings but not on VERITE, where the correlation between external evidence is weaker. 

Nevertheless, it is important to note that employing ``relevant evidence detection'' and training the model to differentiate between relevant and irrelevant evidence ameliorates this issue and significantly improves the performance of \methodname\ on VERITE with all four backbone encoders. 
Notably, by leveraging CLIP ViT L/14, \methodname-SSL strikes the best balance between high performance on both datasets.
These findings strengthen our hypothesis that leveraging multitask learning with both ``verdict prediction' and ``relevant evidence detection'' can improve the overall detection accuracy on real-world misinformation by guiding the network to recognize which pieces of external evidence are actually relevant to support or refute a claim.

Finally, when the \textit{DT} is trained on a larger dataset (``R-NESt + CHASMA-D + NC-t2t'') we observe a reduced but still significant improvement, with \methodname-SSL outperforming DT by 5.78\%. 
Based on this, we deem highly likely that collecting evidence and training \methodname \ on a larger dataset has the potential to further improve performance. 
However, we do not explore this here, since NewsCLIPings+ is currently the largest publicly available dataset for MFC that provides external information as evidence. 

\begin{figure*}
    \centering
    \includegraphics[width=\linewidth]{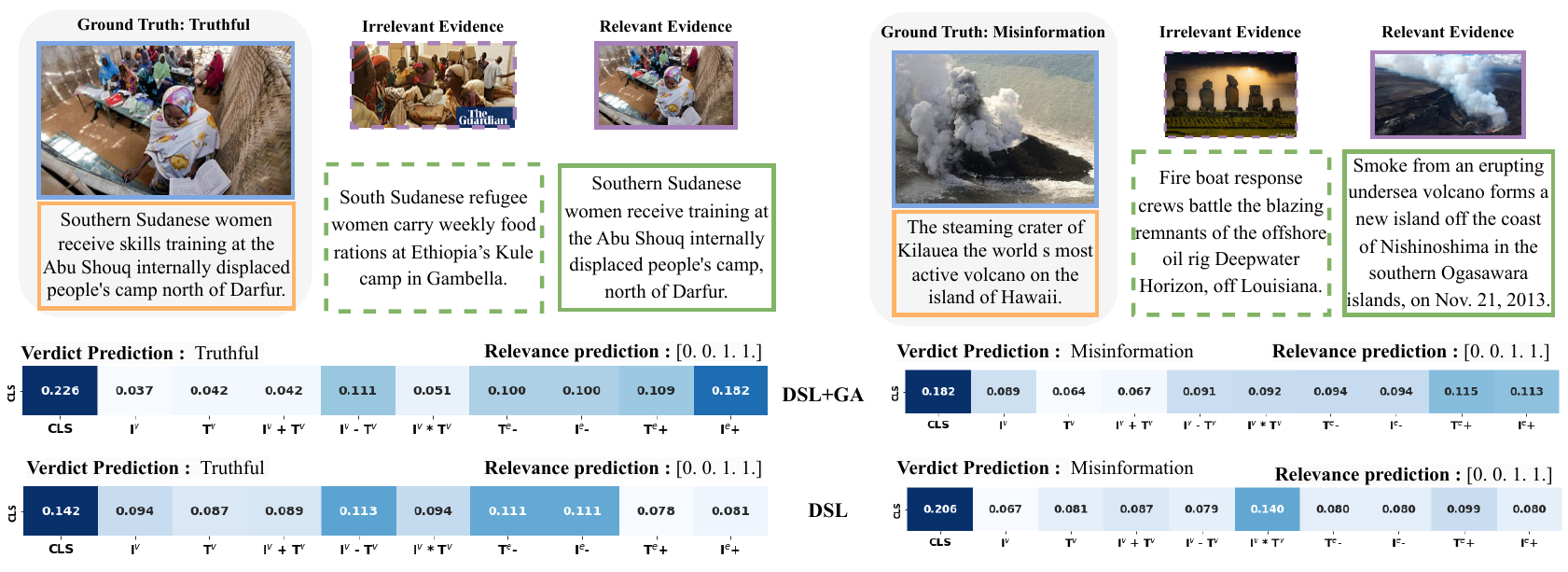}        
    \caption{Inference by \methodname \ variants: DSL and DSL+GA (w/ CLIP ViT B/32) on samples taken from NewsCLIPings+. 
    We report the Attention scores of each method. 
    ``Relevance Ground truth'' is set to [0, 0, 1, 1] for simplicity, regarding [$T^e-$, $I^e-$, $T^e+$, $I^e+$], respectively.
    }

\label{fig:inference}
\end{figure*}

\subsection{Qualitative Analysis}

Figure \ref{fig:inference} depicts the inference process on NewsCLIPings+ test set samples when using two \methodname \ variants, DSL and DSL+GA. 
Focusing on the outcomes, we observe that both models make correct predictions in terms of the overall verdict and evidence relevance.  
In regards to the attention scores, we observe that DSL+GA tends to assign higher attention scores the ``relevant'' items. 
For instance, in the ``Truthful'' pair (left), the evidence retrieval and re-ranking module was able to retrieve the same image $I^e+$ as $I^v+$ and DSL+GA correctly exhibits higher attention scores on $I^e+$ (0.182) followed by $T^e+$ (0.109) 
and lower on irrelevant evidence (0.100). 
On the contrary, the DSL variant shows lower attention scores on $I^e+$ (0.081) and $I^e+$ (0.078) and displays higher attention scores on the irrelevant evidence (0.111).
In the ``Misinformation'' pair, again, DSL+GA assigns higher attention scores to the relevant evidence over irrelevant ones, while DSL only assigns a higher score on $T^e+$ and the same score (0.08) on all other items, including $I^e+$.
By assigning higher attention scores to relevant texts $T^e+$ and even higher to the relevant images $I^e+$, DSL+GA shows the potential to enhance interpretability.

\section{Conclusions}
\label{sec:conclusions}

In this study we address the challenge of evidence-based Multimodal Fact-Checking (MFC) and incorporate Relevant Evidence Detection (RED) as part of the process, where the model, first has to determine which pieces of evidence are relevant to, support or refute, the claim under verification and then proceed to assess its veracity. 
We conduct extensive ablation and comparative experiments to show that the proposed Relevant Evidence Detection Directed Transformer (\methodname) is capable of outperforming its counterparts that are not optimized for RED on VERITE.
Moreover, it outperforms the state-of-the-art on NewsCLIPings+ without requiring numerous evidence, multiple or fine-tuned backbone encoders or additional features.

Despite notable improvements in the field of MFC, it is essential to discuss certain limitations of our research and the field more broadly. 
Following previous studies \cite{abdelnabi2022open, yuan2023support, zhang2023ecenet}, we focused on NewsCLIPings+, which, despite being the only publicly available multimodal dataset with external, non-leaked evidence, it comes with certain constraints. 
Firstly, the dataset consists of a mere 85,000 samples which may not encompass the full spectrum of diversity concerning historical events, prominent figures, and other crucial contextual elements pertinent to misinformation.
Furthermore, NewsCLIPings+ exclusively focuses on ``re-purposed images'' or ``out-of-context misinformation'', omitting categories like ``miscaptioned images'' \cite{papadopoulos2023verite} and only provides short textual evidence -article titles and captions- that may offer limited information in comparison to complete articles.
Therefore, subsequent research endeavors should aim at the collection of more extensive and diverse datasets, that also include full texts as candidate evidence and encompass various forms of multimodal misinformation.

Finally, it is essential to acknowledge that the external information used as ``evidence'' is retrieved from search engines, which may include irrelevant and noisy data and may impact the realization of the full potential of MFC methods. 
Hence, future research should also explore improved methods for gathering, filtering, and assessing the relevance of external information employed as evidence. 
Our work represents a significant first step towards this direction by providing a novel methodological framework for assessing the relevance of external evidence.

\section*{Acknowledgments}
This work is partially funded by the project “vera.ai: VERification Assisted by Artificial Intelligence” under grant agreement no. 101070093.

\bibliographystyle{unsrt}  
\bibliography{manuscript}

\begin{thebibliography}{10}

\bibitem{olan2022fake}
Femi Olan, Uchitha Jayawickrama, Emmanuel~Ogiemwonyi Arakpogun, Jana Suklan, and Shaofeng Liu.
\newblock Fake news on social media: the impact on society.
\newblock {\em Information Systems Frontiers}, pages 1--16, 2022.

\bibitem{bennett2018disinformation}
W~Lance Bennett and Steven Livingston.
\newblock The disinformation order: Disruptive communication and the decline of democratic institutions.
\newblock {\em European journal of communication}, 33(2):122--139, 2018.

\bibitem{cantarella2023does}
Michele Cantarella, Nicol{\`o} Fraccaroli, and Roberto Volpe.
\newblock Does fake news affect voting behaviour?
\newblock {\em Research Policy}, 52(1):104628, 2023.

\bibitem{roozenbeek2020susceptibility}
Jon Roozenbeek, Claudia~R Schneider, Sarah Dryhurst, John Kerr, Alexandra~LJ Freeman, Gabriel Recchia, Anne~Marthe Van Der~Bles, and Sander Van Der~Linden.
\newblock Susceptibility to misinformation about covid-19 around the world.
\newblock {\em Royal Society open science}, 7(10):201199, 2020.

\bibitem{do2022infodemics}
Israel Junior~Borges Do~Nascimento, Ana~Beatriz Pizarro, Jussara~M Almeida, Natasha Azzopardi-Muscat, Marcos~Andr{\'e} Gon{\c{c}}alves, Maria Bj{\"o}rklund, and David Novillo-Ortiz.
\newblock Infodemics and health misinformation: a systematic review of reviews.
\newblock {\em Bulletin of the World Health Organization}, 100(9):544, 2022.

\bibitem{gamir2021multimodal}
Jos{\'e} Gamir-R{\'\i}os, Raquel Tarullo, Miguel Ib{\'a}{\~n}ez-Cuquerella, et~al.
\newblock Multimodal disinformation about otherness on the internet. the spread of racist, xenophobic and islamophobic fake news in 2020.
\newblock {\em An{\`a}lisi}, pages 49--64, 2021.

\bibitem{ruokolainen2020conceptualising}
Hilda Ruokolainen and Gunilla Wid{\'e}n.
\newblock Conceptualising misinformation in the context of asylum seekers.
\newblock {\em Information Processing \& Management}, 57(3):102127, 2020.

\bibitem{zhou2023synthetic}
Jiawei Zhou, Yixuan Zhang, Qianni Luo, Andrea~G Parker, and Munmun De~Choudhury.
\newblock Synthetic lies: Understanding ai-generated misinformation and evaluating algorithmic and human solutions.
\newblock In {\em Proceedings of the 2023 CHI Conference on Human Factors in Computing Systems}, pages 1--20, 2023.

\bibitem{xu2023combating}
Danni Xu, Shaojing Fan, and Mohan Kankanhalli.
\newblock Combating misinformation in the era of generative ai models.
\newblock In {\em Proceedings of the 31st ACM International Conference on Multimedia}, pages 9291--9298, 2023.

\bibitem{aneja2023cosmos}
Shivangi Aneja, Chris Bregler, and Matthias Nie{\ss}ner.
\newblock Cosmos: Catching out-of-context image misuse using self-supervised learning.
\newblock In {\em Proceedings of the AAAI Conference on Artificial Intelligence}, volume~37, pages 14084--14092, 2023.

\bibitem{luo2021newsclippings}
Grace Luo, Trevor Darrell, and Anna Rohrbach.
\newblock Newsclippings: Automatic generation of out-of-context multimodal media.
\newblock In {\em Proceedings of the 2021 Conference on Empirical Methods in Natural Language Processing}, pages 6801--6817, 2021.

\bibitem{hangloo2022combating}
Sakshini Hangloo and Bhavna Arora.
\newblock Combating multimodal fake news on social media: methods, datasets, and future perspective.
\newblock {\em Multimedia systems}, 28(6):2391--2422, 2022.

\bibitem{boididou2018verifying}
Christina Boididou, Stuart~E Middleton, Zhiwei Jin, Symeon Papadopoulos, Duc-Tien Dang-Nguyen, Giulia Boato, and Yiannis Kompatsiaris.
\newblock Verifying information with multimedia content on twitter: a comparative study of automated approaches.
\newblock {\em Multimedia tools and applications}, 77:15545--15571, 2018.

\bibitem{nakamura2020fakeddit}
Kai Nakamura, Sharon Levy, and William~Yang Wang.
\newblock Fakeddit: A new multimodal benchmark dataset for fine-grained fake news detection.
\newblock In {\em Proceedings of the Twelfth Language Resources and Evaluation Conference}, pages 6149--6157, 2020.

\bibitem{jaiswal2017multimedia}
Ayush Jaiswal, Ekraam Sabir, Wael AbdAlmageed, and Premkumar Natarajan.
\newblock Multimedia semantic integrity assessment using joint embedding of images and text.
\newblock In {\em Proceedings of the 25th ACM international conference on Multimedia}, pages 1465--1471, 2017.

\bibitem{papadopoulos2023synthetic}
Stefanos-Iordanis Papadopoulos, Christos Koutlis, Symeon Papadopoulos, and Panagiotis Petrantonakis.
\newblock Synthetic misinformers: Generating and combating multimodal misinformation.
\newblock In {\em Proceedings of the 2nd ACM International Workshop on Multimedia AI against Disinformation}, pages 36--44, 2023.

\bibitem{biamby2022twitter}
Giscard Biamby, Grace Luo, Trevor Darrell, and Anna Rohrbach.
\newblock Twitter-comms: Detecting climate, covid, and military multimodal misinformation.
\newblock In {\em Proceedings of the 2022 Conference of the North American Chapter of the Association for Computational Linguistics: Human Language Technologies}, pages 1530--1549, 2022.

\bibitem{muller2020multimodal}
Eric M{\"u}ller-Budack, Jonas Theiner, Sebastian Diering, Maximilian Idahl, and Ralph Ewerth.
\newblock Multimodal analytics for real-world news using measures of cross-modal entity consistency.
\newblock In {\em Proceedings of the 2020 International Conference on Multimedia Retrieval}, pages 16--25, 2020.

\bibitem{khattar2019mvae}
Dhruv Khattar, Jaipal~Singh Goud, Manish Gupta, and Vasudeva Varma.
\newblock Mvae: Multimodal variational autoencoder for fake news detection.
\newblock In {\em The world wide web conference}, pages 2915--2921, 2019.

\bibitem{wang2018eann}
Yaqing Wang, Fenglong Ma, Zhiwei Jin, Ye~Yuan, Guangxu Xun, Kishlay Jha, Lu~Su, and Jing Gao.
\newblock Eann: Event adversarial neural networks for multi-modal fake news detection.
\newblock In {\em Proceedings of the 24th acm sigkdd international conference on knowledge discovery \& data mining}, pages 849--857, 2018.

\bibitem{yu2022bcmf}
Chuanming Yu, Yinxue Ma, Lu~An, and Gang Li.
\newblock Bcmf: A bidirectional cross-modal fusion model for fake news detection.
\newblock {\em Information Processing \& Management}, 59(5):103063, 2022.

\bibitem{singhal2019spotfake}
Shivangi Singhal, Rajiv~Ratn Shah, Tanmoy Chakraborty, Ponnurangam Kumaraguru, and Shin'ichi Satoh.
\newblock Spotfake: A multi-modal framework for fake news detection.
\newblock In {\em 2019 IEEE fifth international conference on multimedia big data (BigMM)}, pages 39--47. IEEE, 2019.

\bibitem{singhal2022leveraging}
Shivangi Singhal, Tanisha Pandey, Saksham Mrig, Rajiv~Ratn Shah, and Ponnurangam Kumaraguru.
\newblock Leveraging intra and inter modality relationship for multimodal fake news detection.
\newblock In {\em Companion Proceedings of the Web Conference 2022}, pages 726--734, 2022.

\bibitem{guo2022survey}
Zhijiang Guo, Michael Schlichtkrull, and Andreas Vlachos.
\newblock A survey on automated fact-checking.
\newblock {\em Transactions of the Association for Computational Linguistics}, 10:178--206, 2022.

\bibitem{wang2017liar}
William~Yang Wang.
\newblock “liar, liar pants on fire”: A new benchmark dataset for fake news detection.
\newblock In {\em Proceedings of the 55th Annual Meeting of the Association for Computational Linguistics (Volume 2: Short Papers)}, pages 422--426, 2017.

\bibitem{augenstein2019multifc}
Isabelle Augenstein, Christina Lioma, Dongsheng Wang, Lucas~Chaves Lima, Casper Hansen, Christian Hansen, and Jakob~Grue Simonsen.
\newblock Multifc: A real-world multi-domain dataset for evidence-based fact checking of claims.
\newblock In {\em Proceedings of the 2019 Conference on Empirical Methods in Natural Language Processing and the 9th International Joint Conference on Natural Language Processing (EMNLP-IJCNLP)}, pages 4685--4697, 2019.

\bibitem{thorne2018fever}
James Thorne, Andreas Vlachos, Christos Christodoulopoulos, and Arpit Mittal.
\newblock Fever: a large-scale dataset for fact extraction and verification.
\newblock In {\em Proceedings of the 2018 Conference of the North American Chapter of the Association for Computational Linguistics: Human Language Technologies, Volume 1 (Long Papers)}, pages 809--819, 2018.

\bibitem{aly2021feverous}
Rami Aly, Zhijiang Guo, Michael Schlichtkrull, James Thorne, Andreas Vlachos, Christos Christodoulopoulos, Oana Cocarascu, and Arpit Mittal.
\newblock Feverous: Fact extraction and verification over unstructured and structured information.
\newblock {\em arXiv preprint arXiv:2106.05707}, 2021.

\bibitem{yao2023end}
Barry~Menglong Yao, Aditya Shah, Lichao Sun, Jin-Hee Cho, and Lifu Huang.
\newblock End-to-end multimodal fact-checking and explanation generation: A challenging dataset and models.
\newblock In {\em Proceedings of the 46th International ACM SIGIR Conference on Research and Development in Information Retrieval}, pages 2733--2743, 2023.

\bibitem{suryavardan2023factify}
S~Suryavardan, Shreyash Mishra, Parth Patwa, Megha Chakraborty, Anku Rani, Aishwarya Reganti, Aman Chadha, Amitava Das, Amit Sheth, Manoj Chinnakotla, et~al.
\newblock Factify 2: A multimodal fake news and satire news dataset.
\newblock {\em arXiv preprint arXiv:2304.03897}, 2023.

\bibitem{abdelnabi2022open}
Sahar Abdelnabi, Rakibul Hasan, and Mario Fritz.
\newblock Open-domain, content-based, multi-modal fact-checking of out-of-context images via online resources.
\newblock In {\em Proceedings of the IEEE/CVF Conference on Computer Vision and Pattern Recognition}, pages 14940--14949, 2022.

\bibitem{glockner2022missing}
Max Glockner, Yufang Hou, and Iryna Gurevych.
\newblock Missing counter-evidence renders nlp fact-checking unrealistic for misinformation.
\newblock In {\em Proceedings of the 2022 Conference on Empirical Methods in Natural Language Processing}, pages 5916--5936, 2022.

\bibitem{zhang2023ecenet}
Fanrui Zhang, Jiawei Liu, Qiang Zhang, Esther Sun, Jingyi Xie, and Zheng-Jun Zha.
\newblock Ecenet: Explainable and context-enhanced network for muti-modal fact verification.
\newblock In {\em Proceedings of the 31st ACM International Conference on Multimedia}, pages 1231--1240, 2023.

\bibitem{yuan2023support}
Xin Yuan, Jie Guo, Weidong Qiu, Zheng Huang, and Shujun Li.
\newblock Support or refute: Analyzing the stance of evidence to detect out-of-context mis-and disinformation.
\newblock {\em arXiv preprint arXiv:2311.01766}, 2023.

\bibitem{radford2021learning}
Alec Radford, Jong~Wook Kim, Chris Hallacy, Aditya Ramesh, Gabriel Goh, Sandhini Agarwal, Girish Sastry, Amanda Askell, Pamela Mishkin, Jack Clark, et~al.
\newblock Learning transferable visual models from natural language supervision.
\newblock In {\em International conference on machine learning}, pages 8748--8763. PMLR, 2021.

\bibitem{li2021align}
Junnan Li, Ramprasaath Selvaraju, Akhilesh Gotmare, Shafiq Joty, Caiming Xiong, and Steven Chu~Hong Hoi.
\newblock Align before fuse: Vision and language representation learning with momentum distillation.
\newblock {\em Advances in neural information processing systems}, 34:9694--9705, 2021.

\bibitem{li2023blip}
Junnan Li, Dongxu Li, Silvio Savarese, and Steven Hoi.
\newblock Blip-2: Bootstrapping language-image pre-training with frozen image encoders and large language models.
\newblock {\em arXiv preprint arXiv:2301.12597}, 2023.

\bibitem{papadopoulos2023verite}
Stefanos-Iordanis Papadopoulos, Christos Koutlis, Symeon Papadopoulos, and Panagiotis~C Petrantonakis.
\newblock Verite: A robust benchmark for multimodal misinformation detection accounting for unimodal bias.
\newblock {\em arXiv preprint arXiv:2304.14133}, 2023.

\bibitem{mridha2021comprehensive}
Muhammad~F Mridha, Ashfia~Jannat Keya, Md~Abdul Hamid, Muhammad~Mostafa Monowar, and Md~Saifur Rahman.
\newblock A comprehensive review on fake news detection with deep learning.
\newblock {\em IEEE Access}, 9:156151--156170, 2021.

\bibitem{rana2022deepfake}
Md~Shohel Rana, Mohammad~Nur Nobi, Beddhu Murali, and Andrew~H Sung.
\newblock Deepfake detection: A systematic literature review.
\newblock {\em IEEE Access}, 2022.

\bibitem{alam2022survey}
Firoj Alam, Stefano Cresci, Tanmoy Chakraborty, Fabrizio Silvestri, Dimiter Dimitrov, Giovanni Da~San~Martino, Shaden Shaar, Hamed Firooz, Preslav Nakov, et~al.
\newblock A survey on multimodal disinformation detection.
\newblock In {\em Proceedings of the 29th International Conference on Computational Linguistics}, pages 6625--6643. International Committee on Computational Linguistics, 2022.

\bibitem{sabir2018deep}
Ekraam Sabir, Wael AbdAlmageed, Yue Wu, and Prem Natarajan.
\newblock Deep multimodal image-repurposing detection.
\newblock In {\em Proceedings of the 26th ACM international conference on Multimedia}, pages 1337--1345, 2018.

\bibitem{mu2023self}
Michael Mu, Sreyasee Das~Bhattacharjee, and Junsong Yuan.
\newblock Self-supervised distilled learning for multi-modal misinformation identification.
\newblock In {\em Proceedings of the IEEE/CVF Winter Conference on Applications of Computer Vision}, pages 2819--2828, 2023.

\bibitem{suryavardan2023findings}
S~Suryavardan, Shreyash Mishra, Megha Chakraborty, Parth Patwa, Anku Rani, Aman Chadha, Aishwarya Reganti, Amitava Das, Amit Sheth, Manoj Chinnakotla, et~al.
\newblock Findings of factify 2: multimodal fake news detection.
\newblock {\em arXiv preprint arXiv:2307.10475}, 2023.

\bibitem{koutlis2023memefier}
Christos Koutlis, Manos Schinas, and Symeon Papadopoulos.
\newblock Memefier: Dual-stage modality fusion for image meme classification.
\newblock In {\em Proceedings of the 2023 ACM International Conference on Multimedia Retrieval}, pages 586--591, 2023.

\bibitem{kumar2022hate}
Gokul~Karthik Kumar and Karthik Nandakumar.
\newblock Hate-clipper: Multimodal hateful meme classification based on cross-modal interaction of clip features.
\newblock In {\em Proceedings of the Second Workshop on NLP for Positive Impact (NLP4PI)}, pages 171--183, 2022.

\bibitem{koh2021wilds}
Pang~Wei Koh, Shiori Sagawa, Henrik Marklund, Sang~Michael Xie, Marvin Zhang, Akshay Balsubramani, Weihua Hu, Michihiro Yasunaga, Richard~Lanas Phillips, Irena Gao, et~al.
\newblock Wilds: A benchmark of in-the-wild distribution shifts.
\newblock In {\em International Conference on Machine Learning}, pages 5637--5664. PMLR, 2021.

\bibitem{longoni2022news}
Chiara Longoni, Andrey Fradkin, Luca Cian, and Gordon Pennycook.
\newblock News from generative artificial intelligence is believed less.
\newblock In {\em Proceedings of the 2022 ACM Conference on Fairness, Accountability, and Transparency}, pages 97--106, 2022.

\end{thebibliography}

\end{document}